# Resolving the radio loud/radio quiet dichotomy without thick disks

David Garofalo
Department of Physics, Kennesaw State University, Marietta GA 30060, USA

Abstract

Observations of radio loud active galaxies in the *XMM-Newton* archive by Mehdipour & Costantini show a strong anti-correlation between the column density of the ionized wind and the radio loudness parameter, providing evidence that jets may thrive in thin disks. This is in contrast with decades of analytic and numerical work suggesting jet formation is contingent on the presence of an inner, geometrically thick disk structure, which serves to both collimate and accelerate the jet. Thick disks emerge in radiatively inefficient disks which are associated with sub-Eddington as well as super-Eddington accretion regimes yet we show that the inverse correlation between winds and jets survives where it should not, namely in a luminosity regime normally attributed to radio quiet active galaxies which are modeled with thin disks. This along with other lines of evidence argues against thick disks as the foundation behind the radio loud/radio quiet dichotomy, opening up the possibility that jetted versus non jetted black holes may be understood within the context of radiatively efficient thin disk accretion.

1. Introduction

Sikora et al (2007) explored the inverse correlation between radio loudness and Eddington accretion rate for a large sample of active galaxies (AGN) including Seyfert galaxies and LINERs, optically selected quasars, FRI radio galaxies, FRII quasars, and broad line radio galaxies. Despite a clear dichotomy in radio loudness between objects referred to as radio loud and those as radio quiet, the data was insufficiently detailed to shed light on the jet-disk connection near the Eddington limit where we traditionally model the radio quiet subset of the AGN family. This meant that radio loud quasars might be considered different from radio quiet quasars/AGN in the details of the inner accretion disk where radiative efficiency might drop, creating the conditions for thick disk structure that in turn might serve to collimate and accelerate a jet. These ideas have received support from general relativistic simulations over the last decade and a half (Koide et al 1998, 2000; De Villiers & Hawley 2003; De Villiers et al 2003, 2005; Gammie et al 2003; McKinney 2006; Tchekhovskoy et al 2010; McKinney et al 2012). Despite incompatibility with the Soltan argument (Elvis, Risaliti & Zamorani 2002), tension with the observed redshift distribution of radio galaxies and quasars (Garofalo 2013), and with the black hole spin/Meier paradox (Meier 2012; Garofalo et al 2016), hope that such a picture for powerful jet formation would eventually turn out correct has dominated the collective consciousness, in no small part due to the lack of viable alternatives.

Recent observational work on a small group of jetted AGN has confirmed a similar inverse trend as the radio loudness/Eddington ratio of Sikora et al (2007) that instead involves winds and jets (Mehdipour & Costantini 2019). The inverse correlation appears for a number of radio loud

AGN in a narrower range of Eddington ratios compared to Sikora et al (2007). In this paper we select the near-Eddington subset of these radio AGN, where the thin disk assumption most likely applies, to show that the inverse connection between winds and jets still holds. This, we argue, is unexpected and ultimately leads us to conclude that thin versus thick disks cannot constitute the bedrock from which to understand the radio loud/radio quiet dichotomy. Instead, an additional parameter besides accretion rate and black hole spin is needed, which we provide.

2. Discussion

Mehdipour & Costantini (2019) report values of wind column density, radio loudness, and Eddington ratios for 16 X-ray bright radio AGN from *XMM-Newton* whose Eddington ratios $\lambda$ = $L_{bol}/L_{Edd}$ satisfy $0.46 \geq \lambda \geq 0.1$ with $L_{bol}$ the bolometric luminosity and $L_{Edd}$ the Eddington luminosity. They show a significant inverse correlation between the column density and the radio loudness parameter. We extract the subset of these 16 jetted AGN whose Eddington ratios are equal to or greater than 10% and no greater than 46% of the Eddington limit and list them in Table 1 along with the values for the column density and radio loudness. In Figure 1 we plot these objects in the $N_H$-R plane which shows an inverse correlation spanning three decades in both parameters with a linear correlation coefficient of -0.525 which is actually stronger than the -0.395 for the entire dataset.

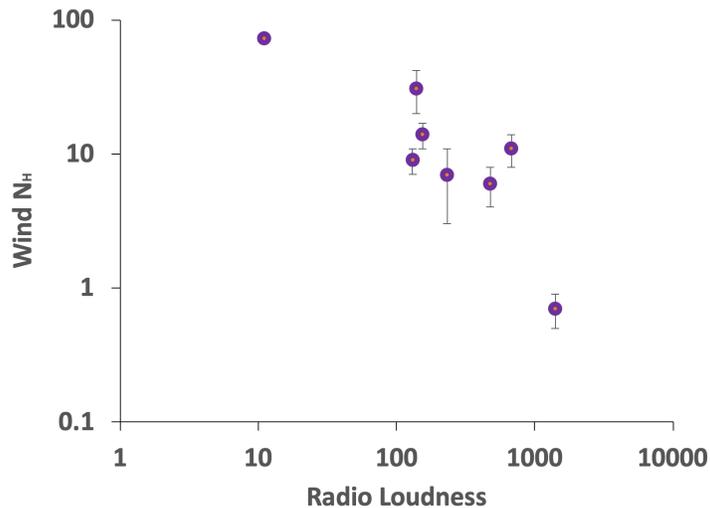

Figure 1: Wind column density (in units of $10^{20}$ cm$^{-2}$) versus radio loudness for the near Eddington sources of Mehdipour & Costantini (2019).

A schematic for understanding our current picture for powerful jet formation in accreting black holes is provided in Figure 2 from Meier (2002). The theoretical boundary between an advection dominated accretion flow (ADAF) and a radiatively efficient accretion flow is $10^{-2}$ of the Eddington accretion so objects that are sufficiently above that boundary but below 1 are thought of both as thin disks and referred to as near Eddington accretors. For accretion disks that accrete at sub-Eddington rates, an ADAF forms in the inner regions near the black hole horizon (left column). When the black hole spin is near zero, no energy is available for extraction from the black hole which leads to a jetless or weak radio source. This models a low ionization nuclear emission line object or LINER (Figure 2 upper left). If the black hole spin is high, powerful jets are generated and a strong radio source or radio galaxy results (Figure 2 lower left). If, on the other hand the disk is thin and radiatively efficient and the black hole spin is near zero, these conditions produce radio quiet AGN (Figure 2 upper right). These low black hole spin, radiatively efficient disks, have been used to model the radio quiet PG quasars and radio quiet Seyfert galaxies as shown in Figure 3 (from Sikora et al 2007). Putting aside some ad-hoc assumptions about tilted disks, the existence of radio quasars and broad line radio galaxies in the context of these ideas requires both high black hole spin and super-Eddington accretion in the inner regions in order to motivate thick inner disks (Figure 2 lower right). Over the ensuing decade since the proposal of these ideas, and despite a small population, the distribution of black hole spin in radio quiet AGN appears to be top heavy (Brenneman 2013). The validity of these spins measurements combined with the models described in Figure 2, implies that radio quasars should have a significant presence in Seyfert galaxies, contrary to observations (Sikora et al 2007). As a result of the possibility of high spinning black holes in many spiral galaxies, a conundrum arises which seems overcome by appealing to near-Eddington accretion rates for radio quiet quasars/AGN, but super Eddington accretion rates in radio loud quasars/AGN. The difference between these two subgroups of AGN would amount to a thick disk geometry in super-Eddington systems that serves to collimate and accelerate jets in the same way as low Eddington accretion systems. Perhaps, it was thought, broad line radio galaxies and FRII quasars are characterized by super-Eddington accretion in their inner regions, reducing their Eddington luminosities but explaining the presence of powerful jets compared to the PG quasars which instead seem to occupy near Eddington luminosities (Figure 3). In summary, then, jets are associated with geometrically thick inner disks that are not present in radio quiet quasars/AGN. We now show how this idea fails.

The above ideas suggest that radio galaxies and radio quasars do not experience near-Eddington luminosity regions which instead are the domain of radio quiet quasars and radio quiet AGN. Since jets should disappear for AGN that live near the Eddington limit, it would not make sense to see both strong jets and the anti-correlation between jets and winds for systems that are best modeled as thin disks. We show, however, that the data in Mehdipour & Costantini (2019) not only provides evidence for powerful jets in AGN that reach their Eddington luminosities, the inverse relation with the wind is just as strong a trend as it is at sub-Eddington values.

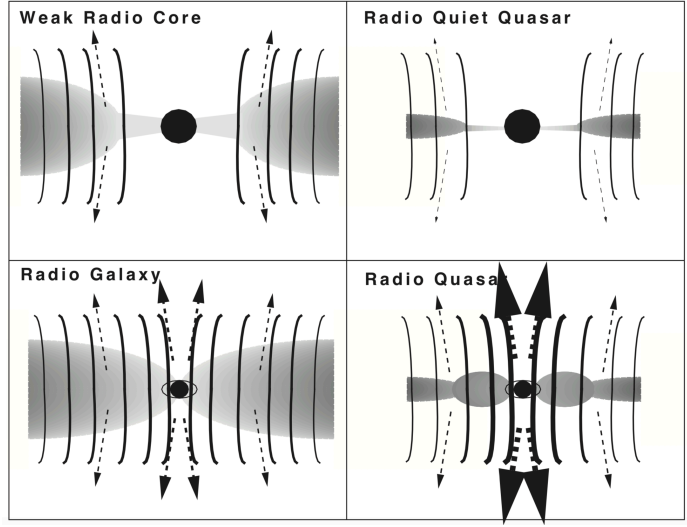

Figure 2: Schematic representation of four possible combinations of accretion state and black hole spin. Top panels depict zero-spin black holes while bottom high prograde spin black holes. Left panels show low accretion rates (ADAF) tori, right panels represent higher accretion rates and thin disk models except for the inner region of the radio quasar (from Meier 2002).

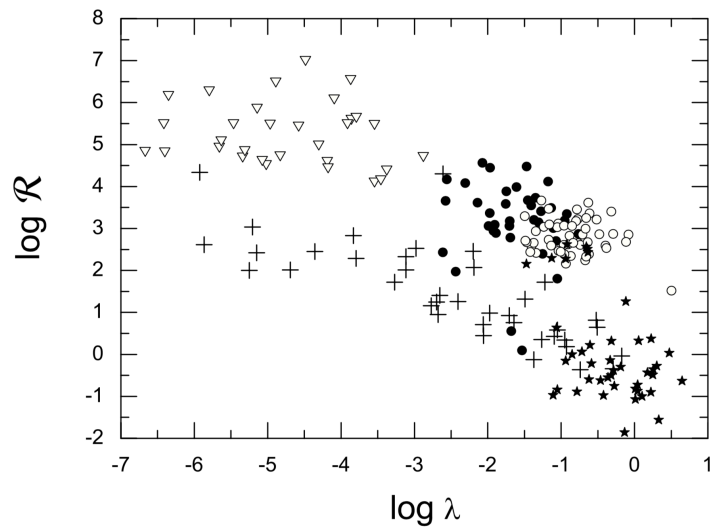

Figure 3: Logarithm of radio loudness versus logarithm of Eddington ratio. Broad line radio galaxies are filled circles, radio loud quasars are open circles, Seyfert galaxies and LINERs are crosses, FRI radio galaxies are open triangles, and PG quasars are filled stars (from Sikora et al 2007 Figure 3).

From the data in Mehdipour & Costantini we pick radio AGN whose Eddington ratios are greater than about 10% (Table 1) and explore whether or not the jet and wind remain correlated. As shown in Figure 1, the radio loudness and the wind column density continue to experience the same trend observed below the Eddington luminosity. Despite the small numbers, the data supports the notion that the jet-disk anti correlation is unrelated to inner disk thickness. Another strong argument against thick disk-based jet engines is that we now have evidence for collimation taking place over vastly larger scales than those of the disk scale height (Nakamura & Asada 2013;

Blandford, Meier & Readhead 2019). The data in Sikora et al (2007) shows the radio loud AGN to be more restrained as they approach the near Eddington limit (i.e. there are no objects with positive log of Eddington ratios unlike the radio quiet ones). The data in this paper argues that the cause of this must be understood in the context of thin disks. In short, the radio loud/radio quiet dichotomy must be understood as the result of a difference in some additional parameter for thin disks around high spinning black holes. This has been done in Garofalo, Evans & Sambruna (2010) with co-rotation of the disk compared to the black hole rotation for radio quiet AGN (the PG quasars of Sikora et al) and counterrotation of the disk for radio loud quasars (the radio quasars and broad line radio galaxies of Sikora et al). According to this picture black hole spin cannot be negligible in order for a powerful jet to exist but a high spinning black hole in co-rotation with its accretion disk is subject to strong winds and jet suppression (Neilsen & Lee 2009; Ponti et al 2012). The strong wind is associated with the large radiative efficiency of the thin disk which in turn is due to the small value of the innermost stable circular orbit (ISCO). In the counterrotating case, instead, the ISCO for a high spinning black hole is almost an order of magnitude large which leads to weaker radiative efficiency (compatible with Figure 3 of Sikora et al 2007) compared to high spin black holes in co-rotation with the disk. As a result of the lower radiative efficiency, the disk wind produces relatively weaker jet suppression which allows the jet to co-exist in a thin disk.

| Source | Wind | R | $\lambda$ |
|---|---|---|---|
| 1H 0323+342 | 9 | 130 | 0.46 |
| 3C 120 | 14 | 154 | 0.27 |
| 3C 273 | 0.7 | 1407 | 0.12 |
| 4C + 31.63 | 11 | 676 | 0.11 |
| 4C + 34.47 | 31 | 139 | 0.41 |
| III Zw 2 | 7 | 233 | 0.21 |
| Mrk 896 | 73 | 11 | 0.1 |
| PKS 0405-12 | 6 | 477 | 0.3 |

Table 1: Near-Eddington sources from Mehdipour & Costantini (2019). Column 1 source name; column 2 wind column density x $10^{20}$; column 3 radio loudness parameter; column 4 Eddington ratio $\lambda$. The sources have Eddington ratios between 10 and 46 percent. In terms of the log of the Eddington ratio this places these objects in between -1 and -0.33 which places them quite close to the log of the Eddington limit in Figure 3 which is at 0.

3. Conclusions

The development of ideas concerning the jet-disk connection for accreting black holes over the last two decades has produced a framework in which sub-Eddington bolometric luminosities are associated with powerful jets as a result of low and high accretion rates producing advection dominated accretion and thick disks. We have shown that observations of powerful jets exist for near-Eddington accretion systems, and that the mechanism behind the jet-disk connection appears to be fully operative in standard thin accretion disks. Because these observations suggest that standard disks inform our models for the radio loud as well as the radio quiet population, and plenty of radio quiet AGN have high measured spin values, there must be an additional parameter that breaks the degeneracy associated with high black hole spin thin disks. In short, we need a reason for high spinning black holes accreting in radiatively efficient disks to produce both radio quiet as well as radio loud AGN. This additional parameter turns out to be disk orientation with counterrotation associated with powerful jets and weaker winds and weaker jets and more powerful winds for corotating disks.

References

Blandford, R., Meier, D., Readhead, A., 2019, ARA&A
De Villiers, J.-P., Hawley, J.F., 2003, ApJ, 589, 458
De Villiers, J.-P. et al 2005, ApJ, 620, 878
Elvis, M., Risaliti, G., Zamorani, G., 2002, ApJ, 565, L75
Gammie, C.F., et al, 2003, ApJ, 589, 444
Garofalo D., Evans D.A., Sambruna, R.M., 2010, MNRAS, 406, 975
Garofalo, D., 2013, AdAst, 213105
Garofalo, D., Kim, M.I., Christian, D.J., Hollingworth, E., Lowery, A., Harmon, M., 2016, ApJ, 817, 170
Koide, S. et al 1998, ApJ, 495, L63
Koide, S. et al, 2000, ApJ, 536, 668
McKinney, J., 2006, MNRAS, 368, 1561
McKinney, J. et al 2012, MNRAS, 423, 3083
Meier , D.L., 2001, ApJ, 548,
Meier, D.L., 2002, New Astronomy Reviews, 46, 247
Mehdipour M. & Costantini, E., 2019, A&A in press
Nakamura, M., Asada, K., 2013, ApJ, 775, 118
Neilsen, J. & Lee, J.C., 2009, Nature, 458, 481
Ponti, G. et al 2012, MNRAS, 422, L11
Sikora, M., Stawarz, L., Lasota, J.-P., 2007, ApJ, 658, 815
Tchekhovskoy, A. et al 2010, ApJ, 711, 50